\documentclass[twocolumn,aps,prb]{revtex4}

\usepackage{graphicx,amsfonts,amssymb,amsmath} 
\usepackage{epsfig}
\usepackage{color}
\usepackage[colorlinks=true,linkcolor=black,citecolor=blue]{hyperref}

\usepackage[normalem]{ulem}

\newcommand{\Ham}{\mathcal{H}}
\newcommand{\beq}[0]{\begin{equation}}
\newcommand{\eeq}[0]{\end{equation}}
\newcommand{\brackets}[1]{\left( #1 \right)}
\newcommand{\pd}[2]{\frac{\partial #1}{\partial #2}} 
\newcommand{\pdd}[2]{\frac{\partial^2 #1}{\partial #2^2}}
\renewcommand{\d}[1]{\, \mathrm{d} #1} 
\renewcommand{\v}[1]{\ensuremath{\mathbf{#1}}} 
\newcommand{\R}{\mathbb{R}}
\newcommand{\avg}[1]{\left< #1 \right>} 

\begin{document}

\title{A reduced model for precessional switching of thin-film nanomagnets under the influence of spin-torque}

\author{Ross G. Lund$^1$, Gabriel D. Chaves-O'Flynn$^2$, Andrew D. Kent$^2$,  Cyrill B. Muratov$^1$}

\affiliation{
$^1$Department of Mathematical Sciences, New Jersey Institute of Technology , University Heights, Newark, NJ 07102, USA\\
$^2$Department of Physics, New York University, 4 Washington Place, New York, NY 10003, USA
}

\date{\today}

\begin{abstract}
  We study the magnetization dynamics of thin-film magnetic elements
  with in-plane magnetization
    subject to a spin-current flowing perpendicular to the
  film plane. We derive a reduced partial differential
  equation for the in-plane magnetization
  angle in a weakly damped regime. We then apply this model to study
  the experimentally relevant problem of switching of an elliptical
  element when the spin-polarization has a component perpendicular to
  the film plane,
  restricting the reduced model to a macrospin approximation. The
  macrospin ordinary differential equation is treated analytically as
  a weakly damped Hamiltonian system, and an orbit-averaging method is
  used to understand transitions in solution behaviors in terms of a
  discrete dynamical system. The predictions of our reduced model are
  compared to those of the full Landau--Lifshitz--Gilbert--Slonczewski
  equation for a macrospin.  
  \end{abstract}


\maketitle

\section{Introduction}
Magnetization dynamics in the presence of spin-transfer torques is a
very active area of research with applications to magnetic memory
devices and
oscillators~\cite{BaderParkin2010,BrataasKentOhno2012,KentWorledge2015}.
Some basic questions relate to the types of magnetization dynamics
that can be excited and the time scales on which the dynamics occurs.
Many of the experimental studies of spin-transfer torques are on thin
film magnetic elements patterned into asymmetric shapes (e.g. an
ellipse) in which the demagnetizing field strongly confines the
magnetization to the film plane. Analytic models that capture the
resulting nearly in-plane magnetization dynamics (see e.g.
\cite{GarciaCerveraE01, DKMO, KohnSlas05Dynamics, MuratovOsipov06,
  CapellaOttoMelcher}) can lead to new insights and guide experimental
studies and device design. A macrospin model that treats the entire
magnetization of the element as a single vector of fixed length is a
starting point for most analyses.

The focus of this paper is on a thin-film magnetic element excited by
a spin-polarized current that has an out-of-plane component. This
out-of-plane component of spin-polarization can lead to magnetization
precession about the film normal or magnetization reversal. The
former dynamics would be desired for a spin-transfer torque
oscillator, while the latter dynamics would be essential in a magnetic
memory device. A device in which a perpendicular component of
spin-polarization is applied to an in-plane magnetized element was
proposed in {Ref.~[\onlinecite{Kent2004}]} and has been studied
experimentally {\cite{Liu2010, Liu2012, Ye2015}}. There have also
been a number of models that have considered the influence of thermal
noise on the resulting dynamics, e.g., {on} the rate of switching
and the dephasing of the oscillator motion
\cite{Newhall2013,Pinna2013,Pinna2014}.

Here we consider a weakly damped asymptotic regime of the
Landau--Lifshitz--Gilbert--Slonczewski (LLGS) equation for a thin-film
ferromagnet, in which the oscillatory nature of the in-plane dynamics
is highlighted. 
In this regime, we derive a reduced partial
differential equation (PDE) for the in-plane
magnetization dynamics under applied spin-torque, which is a
generalization of the underdamped wave-like model due to Capella,
Melcher and Otto \cite{CapellaOttoMelcher}. We then analyze the
  solutions of this equation under the macrospin (spatially uniform)
  approximation, and discuss the predictions of such a model in the
context of previous numerical studies of the full LLGS equation
\cite{ChavesKent15}.

The rest of this article is organized as follows. In Sec. II, we
perform an asymptotic derivation of the reduced underdamped equation
for the in-plane magnetization dynamics in a thin-film element of
arbitrary cross section, by first making a thin-film approximation to
the LLGS equation, then a weak-damping approximation. In Sec. III, we
then further reduce to a macrospin ordinary differential equation
(ODE) by spatial averaging of the underdamped PDE, and restrict to the
particular case of a soft elliptical element. A brief parametric study
of the ODE solutions is then presented, varying the spin-current
parameters. In Sec. IV, we make an analytical study of the macrospin
equation using an orbit-averaging method to reduce to a discrete
dynamical system, and compare its predictions to the full ODE
solutions. In Sec. V, we seek to understand transitions between the
different solution trajectories (and thus predict current-parameter
values when the system will either switch or precess) by studying the
discrete dynamical system derived in Sec. IV. Finally, we summarize
our findings in Sec. VI.

\section{Reduced model}
We consider a domain $\Omega\subset\R^3$ occupied by a ferromagnetic
film with cross-section $D\subset \R^2$ and thickness $d$, i.e.,
  $\Omega = D \times (0, d)$.
Under the influence of a spin-polarized electric current applied perpendicular to the film plane, the magnetization vector $\v{m}=\v{m}(\v{r},t)$, with $|\v{m}|=1$ in $\Omega$ and $0$ outside, satisfies the LLGS equation (in SI units)
\beq
\pd{\v{m}}{t} = - \gamma\mu_0\v{m} \times \v{H}_{\text{eff}} + {\alpha} \v{m} \times \pd{\v{m}}{t} + {\tau_{\text{STT}}}
\eeq
in $\Omega$, with ${\partial \v{m}}/{\partial n}=(\v{n}\cdot\nabla)\v{m}=0$ on $\partial \Omega$, where $\v{n}$ is the outward unit normal to $\partial \Omega$.
In the above, $\alpha > 0$ is the Gilbert damping parameter, $\gamma$ is the gyromagnetic ratio, $\mu_0$ is the permeability of free space,
$
\v{H}_{\text{eff}} = -\frac{1}{\mu_0M_s}\frac{\delta E}{\delta \v{m}}
$
is the effective magnetic field,
\begin{multline}
E(\v{m}) = \int_\Omega \Big(A |\nabla \v{m}|^2 + K \Phi(\v{m}) - \mu_0 M_s \v{H}_{\text{ext}}\cdot\v{m}) \Big)\d^3 r\\
+ \mu_0 M_s^2 \int_{\R^3}\int_{\R^3} \frac{\nabla \cdot \v{m}(\v{r})\nabla \cdot \v{m}(\v{r'})}{8\pi |\v{r}-\v{r}'|}\d^3 r \d^3 r'
\end{multline}
is the micromagnetic energy with exchange constant $A$, anisotropy
constant $K$, crystalline anisotropy function $\Phi$, external
magnetic field $\v{H}_{\text{ext}}$, and saturation magnetization
$M_s$. Additionally, the Slonczewski spin-transfer torque
$\tau_\text{STT}$ is given by \beq \tau_\text{STT} = -\frac{\eta
  \gamma \hbar j}{2 d e M_s} \v{m} \times\v{m} \times \v{p}, \eeq
where $j$ is the density of current passing perpendicularly through
the film, $e$ is the elementary charge (positive), $\v{p}$ is the
spin-polarization direction, and $\eta \in(0,1]$ is the
spin-polarization efficiency.

We now seek to nondimensionalize the above system. Let
\beq
\ell = \sqrt{\frac{2A}{\mu_0M_s^2}}, \quad Q = \frac{2K}{\mu_0M_s^2}, \quad \v{h}_{\text{ext}} = \frac{\v{H}_{\text{ext}}}{ M_s}.
\eeq
We then rescale space and time as
\beq
\v{r} \to \ell \v{r}, \quad t \to \frac{t}{\gamma \mu_0 M_s} ,
\eeq
obtaining the nondimensional form 
\beq
\pd{\v{m}}{t} = -\v{m} \times \v{h}_{\text{eff}} + \alpha \v{m} \times \pd{\v{m}}{t} - \beta \v{m} \times\v{m} \times \v{p},
\label{LLG}
\eeq where $\v{h}_{\text{eff}} = \v{H}_{\text{eff}}/M_s$, and \beq
\beta = \frac{\eta \hbar j}{2 d e \mu_0M_s^2} \eeq is the
dimensionless spin-torque strength.

Since we are interested in thin films, we now assume that $\v{m}$ is
independent of the film thickness. Then, after rescaling 
\beq
E\to \mu_0 M_s^2 d\ell^2 E,
\eeq
we have $\v{h}_{\text{eff}} \simeq -\frac{\delta E}{\delta \v{m}}$,
where $E$ is given by a local energy functional defined on the
  (rescaled) two-dimensional domain $D$ (see, e.g., Ref. [\onlinecite{KohnSlas05Gamma}]):
\begin{multline}
E(\v{m}) \simeq \frac12 \int_D \left(|\nabla \v{m}|^2 + Q \Phi(\v{m}) - 2\v{h}_{\text{ext}} \cdot \v{m}\right) \d^2 r \\
+ \frac12 \int_D m_\perp^2 \d^2 r + \frac{1}{4\pi}\delta |\ln \lambda| \int_{\partial D} (\v{m} \cdot \v{n})^2 \d s,
\label{ThinFilmEnergy}
\end{multline}
in which now $\v{m}:D\to \mathbb{S}^2$, $m_\perp$ is its out-of-plane
component, $\delta = d/\ell$ is the dimensionless film thickness, and
$\lambda = d /L \ll 1 $ (where $L$ is the lateral size of the film) is the
  film's aspect ratio. The effective field is given explicitly by
\beq
\v{h}_{\text{eff}} = \Delta \v{m} - \frac{Q}{2} \nabla_{\v{m}}\Phi(\v{m}) - m_\perp \v{e}_z + \v{h}_{\text{ext}},
\label{field}
\eeq 
and $\v{m}$ satisfies equation \eqref{LLG} in $D$ with the boundary condition
\beq
\pd{\v{m}}{n} = -\frac{1}{2\pi}\delta |\ln \lambda| (\v{m} \cdot
\v{n})(\v{n} - ( \v{m}\cdot \v{n}) \, \v{m} )
\label{BCm}
\eeq on $\partial D$.

We now parametrize $\v{m}$ in terms of spherical angles as
\beq
\v{m}=(-\sin\theta \cos \phi, \cos\theta \cos \phi, \sin \phi),
\label{Mparam}
\eeq
and the current polarization direction $\v{p}$ in terms of an in-plane angle $\psi$ and its out-of-plane component $p_\perp$ as
\beq
\v{p} = \frac{1}{\sqrt{1+p_\perp^2}}(-\sin\psi, \cos \psi, p_\perp).
\eeq
Writing ${\beta}_* = {\beta}/\sqrt{1+p_\perp^2}$, after some algebra, one may then write equation \eqref{LLG} as the system
\begin{multline}
\pd{\phi}{t} = -\frac{1}{\cos\phi} \v{h}_{\text{eff}}\cdot \v{m}_\theta + \alpha \cos \phi \pd{\theta}{t} \\+ {\beta}_*(p_\perp \cos \phi - \sin\phi\cos(\theta-\psi)),
\end{multline}
\beq
-\cos\phi\pd{\theta}{t} = -\v{h}_{\text{eff}}\cdot \v{m}_\phi + \alpha \pd{\phi}{t} + {\beta}_*\sin(\theta-\psi),\eeq
where $\v{m}_\theta = \partial \v{m}/\partial \theta$ and $\v{m}_\phi = \partial \v{m}/\partial \phi$ for $\v{m}$ given by \eqref{Mparam}.
Again, since we are working in a soft thin film, we assume $\phi \ll 1$ and that the out-of-plane component of the effective field in equation \eqref{field} is dominated by the term $\v{h}_\text{eff} \cdot \v{e}_z \simeq - m_\perp = -\sin \phi$. Note that this assumes that the crystalline anisotropy and external field terms in the out-of-plane directions are relatively small, so we assume the external field is only in plane, though it is still possible to include a perpendicular anisotropy simply by renormalizing the constant in front of the $m_\perp$ term in $\v{h}_\text{eff}$. We then linearize the above system in $\phi$, yielding 
\beq
\pd{\phi}{t} = \frac{\delta \mathcal{E}}{\delta \theta} + \alpha \pd{\theta}{t} + {\beta}_*(p_\perp - \phi \cos(\theta-\psi)),\eeq
\begin{multline} -\pd{\theta}{t} = \phi + {\beta}_*\sin(\theta-\psi)\\+ \phi(-h_x \sin\theta +  h_y \cos\theta)   + \alpha \pd{\phi}{t} .
\label{LLGlin2}
\end{multline}
where $h_x = \v{h}_{\text{eff}}\cdot\v{e}_x$ and $h_y = \v{h}_{\text{eff}}\cdot\v{e}_y$, and $\mathcal{E}(\theta)$ is $E(\v{m})$ evaluated at $\phi = 0$.

We now note that the last two terms in \eqref{LLGlin2} are negligible relative to $\phi$ whenever $|h_x|, |h_y|$ and $\alpha$ are small, which is true of typical clean thin-film samples of sufficiently large lateral extent. Neglecting these terms, one has
\begin{align}
\pd{\phi}{t} &= \frac{\delta \mathcal{E}}{\delta \theta} + \alpha \pd{\theta}{t} + \beta_*(p_\perp - \phi \cos(\theta-\psi)),\label{LLGunscaled1}\\
-\pd{\theta}{t} &= \beta_*\sin(\theta-\psi)+\phi.
\label{LLGunscaled2}
\end{align}
Then, differentiating \eqref{LLGunscaled2} with respect to $t$ and using the result along with \eqref{LLGunscaled2} to eliminate $\phi$ and $\pd{\phi}{t}$ from \eqref{LLGunscaled1}, we find a second-order in time equation for $\theta$:
\begin{multline}
0 = \pdd{\theta}{t} + \pd{\theta}{t}(\alpha + 2 \beta_*\cos(\theta-\psi)) +\frac{\delta \mathcal{E}}{\delta \theta} \\+ \beta_* p_\perp + \beta_*^2 \sin(\theta - \psi)\cos(\theta-\psi),
\label{reduced}
\end{multline}
where, explicitly, one has
\beq
\frac{\delta \mathcal{E}}{\delta \theta} = -\Delta \theta + \frac{Q}{2} {\tilde{\Phi}'}(\theta) + \v{h}_{\text{ext}}\cdot(\cos \theta,\sin \theta),
\eeq
and $\tilde{\Phi}(\theta) = \Phi(\v{m}(\theta))$.
In turn, from the boundary condition on $\v{m}$ in \eqref{BCm}, we can derive the boundary condition for $\theta$ as
\beq
\v{n}\cdot \nabla \theta = \frac{1}{2\pi}\delta|\ln \lambda|\sin(\theta - \varphi) \cos(\theta - \varphi),
\label{BCtheta}
\eeq
where $\varphi$ is the angle parametrizing the normal $\v{n}$ to
$\partial D$ via $\v{n} = (-\sin\varphi,\cos\varphi)$.

The model comprised of \eqref{reduced}--\eqref{BCtheta} is a
damped-driven wave-like PDE for $\theta$, which coincides with
the reduced model of Ref.~[\onlinecite{CapellaOttoMelcher}] for
vanishing spin-current density in an infinite sample. This constitutes
our  reduced PDE model for magnetization dynamics in
thin-film elements under the influence of out-of-plane spin currents.
It is easy to see that all of the terms in \eqref{reduced}
balance when the parameters are chosen so as to satisfy
 \beq \beta_* \sim p_\perp \sim \alpha \sim
{Q}^{1/2} \sim | \v{h}_{\text{ext}}|^{1/2} \sim
\frac{\ell}{L}\sim \delta|\ln \lambda|. \label{scalepde} \eeq
This shows that
  it should be possible to rigorously obtain the reduced model in
  \eqref{reduced}--\eqref{BCtheta} in the asymptotic limit of $L \to
  \infty$ and $\alpha, \beta_*, p_\perp, Q, |\mathbf h_\mathrm{ext}|,
  \delta \to 0$ jointly, so that \eqref{scalepde} holds.

\section{Macrospin switching}
In this section we study the behavior of the reduced model
\eqref{reduced}--\eqref{BCtheta} in the approximation that the
magnetization is spatially uniform on an elliptical domain, and
compare the solution phenomenology to that found by simulating the
LLGS equation in the same physical situation, as studied in
Ref.~[\onlinecite{ChavesKent15}].

\subsection{Derivation of macrospin model} Integrating equation
\eqref{reduced} over the domain $D$ and using the boundary condition
\eqref{BCtheta}, we have \begin{multline} \int_D \left(\pdd{\theta}{t}
    + \pd{\theta}{t}(\alpha + 2 \beta_*\cos(\theta-\psi))
  \right. \\\left.+ \beta_* p_\perp+ \beta_*^2 \sin(\theta -
    \psi)\cos(\theta-\psi)\right. \\\left.+ \frac{Q}{2}
    {\tilde{\Phi}'}(\theta) + \v{h}_{\text{ext}}\cdot(\cos \theta,\sin
    \theta) \right) \d^2 r \\= \frac{1}{2\pi}\delta|\ln \lambda|
  \int_{\partial D} \sin(\theta - \varphi) \cos(\theta - \varphi) \d
  s. \end{multline} Assume now that $\theta$ does not vary appreciably
across the domain $D$, which makes sense in magnetic elements that are
not too large. This allows us to replace $\theta(\v{r},t)$ by its
spatial average
$\bar{\theta}(t) = \frac{1}{|D|}\int_D\theta(\v{r},t) \d^2 r$, where
$|D|$ stands for the area of $D$ in the units of $\ell^2$.  Denoting
time derivatives by overdots, and omitting the bar on $\bar{\theta}$
for notational simplicity, this spatial averaging leads to the
following ODE for $\theta(t)$:
\begin{multline} \ddot{\theta}
  +\dot{\theta}\brackets{\alpha + 2\beta_* \cos(\theta-\psi)} + 
  \beta_*^2 \sin(\theta - \psi)\cos(\theta-\psi) \\+\beta_*
  p_\perp + \frac{Q}{2} {\tilde{\Phi}'}(\theta) +
  \v{h}_{\text{ext}}\cdot(\cos \theta,\sin \theta) \\=
  \frac{\delta|\ln \lambda|}{4\pi |D|} \, \sin 2 \theta
    \int_{\partial D} \cos (2 \varphi) \d s \\ - \frac{\delta|\ln
      \lambda|}{4\pi |D|} \, \cos 2 \theta 
      \int_{\partial D} \sin (2 \varphi) \d s .
  \label{MacrospinGeneral} \end{multline}

\begin{figure*} \begin{center}
    \includegraphics[width=.9\textwidth]{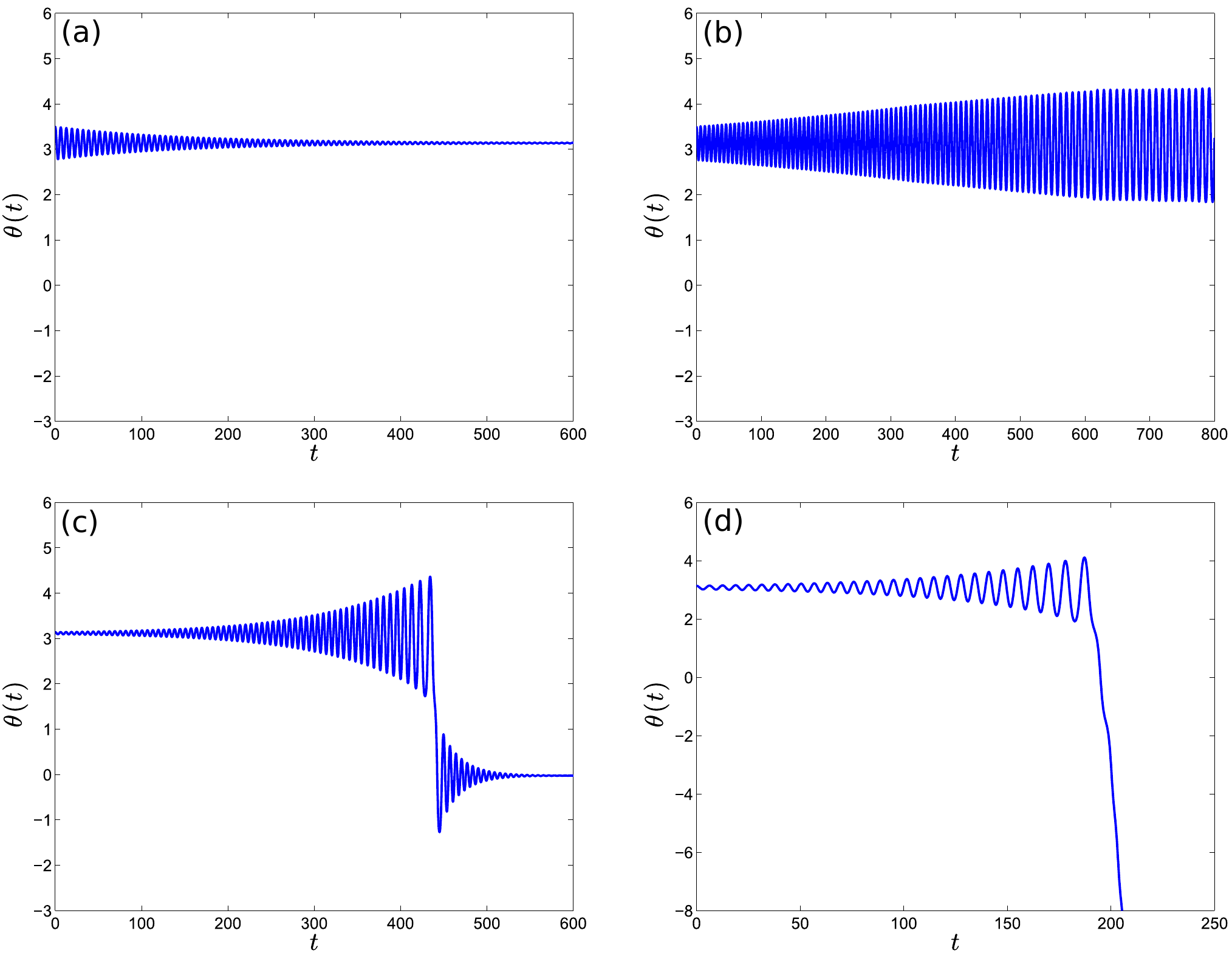}
    \caption{Solutions of macrospin equation \eqref{reducedODE} for
      $\alpha=0.01$, $\Lambda=0.1$. In (a), $p_\perp=0.2$, $\sigma =0.03$:
      decaying solution; in (b), $p_\perp=0.2$, $\sigma =0.06$:
      limit cycle solution (the initial conditions in (a) and
      (b) are $\theta(0)=3.5$, to better visualize the behavior).
      In (c), $p_\perp=0.3$, $\sigma =0.08$: switching solution; in
      (d), $p_\perp=0.6$, $\sigma =0.1$: precessing solution. }
\label{odesols}
\end{center}
\end{figure*}

Next, we consider a particular physical situation in which to study
the macrospin equation, motivated by previous work \cite{Liu2010,
  Liu2012}. As in Refs.~[\onlinecite{Pinna2013, Pinna2014,
  ChavesKent15}], we consider an elliptical thin-film element (recall
that lengths are now measured in the units of $\ell$):
  \begin{align}
    \label{eq:1}
    D = \left\{ (x, y) \ : \  {x^2
    \over a^2} + {y^2 \over b^2} < 1 \right\},
  \end{align}
with no in-plane crystalline anisotropy, $Q=0$, and no external
field, $\v{h}_{\text{ext}}=0$. We take the long axis of the ellipse to
be aligned with the $\v{e}_y$-direction, i.e. $b > a$, with the
in-plane component of current polarization also aligned along this
direction, i.e.,
  taking $\psi =0$. One can then  compute the integral over the boundary in
equation  \eqref{MacrospinGeneral} explicitly, leading to the equation
\begin{multline}
  \ddot{\theta} +\dot{\theta}\brackets{\alpha + \beta_*
  \cos\theta} + \Lambda  \sin \theta\cos\theta \\+ \beta_*^2 \sin\theta\cos\theta+\beta_* p_\perp  = 0,
\end{multline}
where we introduced the geometric parameter $0 < \Lambda \ll 1$
obtained by an explicit integration:
\beq
\Lambda = \frac{\delta|\ln \lambda|}{2\pi^2
  ab} \int_0^{2\pi} \frac{b^2 \cos^2 \tau - a^2\sin^2 \tau}{\sqrt{b^2 \cos^2
    \tau + a^2\sin^2 \tau}} \d \tau. \eeq This may be computed in
terms of elliptic integrals, though the expression is cumbersome so we
omit it here. Importantly, up to a factor depending only on the
  eccentricity the value of $\Lambda$ is given by 
\beq 
\Lambda \sim \frac{d}{L} \ln \frac{L}{d}. 
\eeq 
For example, for an
  elliptical nanomagnet with dimensions $100 \times 30 \times 2.5$ nm
  (similar to those considered in Ref.~[\onlinecite{ChavesKent15}]),
  this yields $\Lambda \simeq 0.1$.

  It is convenient to rescale time by $\sqrt{\Lambda}$ and divide
  through by $\Lambda$, yielding
\begin{multline}
 \ddot{\theta}  +
\frac{1}{\sqrt{\Lambda}}\dot{\theta}\brackets{\alpha + 2\sigma \Lambda \cos\theta} + \sin\theta \cos \theta \\+ \sigma p_\perp+\sigma^2 \Lambda \sin\theta \cos\theta = 0,
\label{reducedODE} 
\end{multline}
where we introduced $\sigma = \beta_*/\Lambda$.  We then apply this
ODE to model the problem of switching of the thin-film elements,
taking the initial in-plane magnetization direction to be static and
aligned along the easy axis, antiparallel to the in-plane component of
the spin-current polarization. Thus, we take \beq \theta(0)=\pi,\quad
\dot{\theta}(0) = 0,
\label{ICMacro}
\eeq
and study the resulting initial value problem. 

\subsection{Solution phenomenology}

Let us briefly investigate the solution phenomenology as the
dimensionless spin-current parameters $\sigma$ and $p_\perp$ are
varied, with the material parameters, $\alpha$ and $\Lambda$, fixed.
We take all parameters to be constant in time for simplicity. We find,
by numerical integration, 4 types of solution to the initial value
problem defined above. The sample solution curves are displayed in
Fig. \ref{odesols} below. The first (panel (a)) occurs for small
values of $\sigma$, and consists simply of oscillations of $\theta$
around a fixed point close to the long axis of the ellipse, which
decay in amplitude towards the fixed point, without switching.

Secondly (panel (b)), still below the switching threshold, the same
oscillations about the fixed point can reach a finite fixed amplitude
and persist without switching. This behavior corresponds to the
  onset of relatively small amplitude
  limit-cycle oscillations around the fixed point.

Thirdly (panel (c)), increasing either $\sigma, p_\perp$ or both, we obtain switching solutions. These have initial oscillations in $\theta$ about the fixed point near $\pi$, which increase in amplitude, and eventually cross the short axis of the ellipse at $\theta=\pi/2$. Then $\theta$ oscillates about the fixed point near 0, and the oscillations decay in amplitude toward the fixed point.

Finally (panel (d)), further increasing $\sigma$ and $p_\perp$ we obtain precessing solutions. Here, the initial oscillations about the fixed point near $\pi$ quickly grow to cross $\pi/2$, after which $\theta$ continues to decrease for all $t$, the magnetization making full precessions around the out-of-plane axis.

\section{Half-period orbit-averaging approach}
We now seek to gain some analytical insight into the transitions
between the solution types discussed above. We do this by averaging
over half-periods of the oscillations observed in the solutions to
generate a discrete dynamical system which describes the evolution of
the energy of a solution $\theta(t)$ on half-period time intervals.

Firstly, we observe that in the relevant parameter regimes the reduced
equation \eqref{reducedODE} can be seen as a weakly perturbed
Hamiltonian system. We consider both $\alpha$ and $\Lambda$ small,
with $\alpha \lesssim \sqrt{\Lambda}$, and assume
$\sigma \sim \alpha/\Lambda$ and $\sigma p_\perp \lesssim 1$.  The
arguments below can be rigorously justified by considering, for
example, the limit $\Lambda \to 0$ while assuming that
$\alpha = O(\Lambda)$ and that the values of $\sigma$ and $p_\perp$
are fixed. This limit may be achieved in the original model by sending
jointly $d \to 0$ and $L \to \infty$, while
keeping\cite{KohnSlas05Gamma}
  \begin{align}
    {L d \over \ell^2} \ln {L \over d} \lesssim 1.
  \end{align}
  The last condition ensures the consistency of the assumption that
  $\theta$ does not vary appreciably throughout $D$.

Introducing 
$\omega(t) = \dot{\theta}(t)$, 
\eqref{reducedODE}  can be written to leading order as
\beq
\dot{\theta} = \pd{\Ham}{\omega}, \quad \dot{\omega} = -
\pd{\Ham}{\theta}, 
\label{HAMILTONIANSYSTEM}
\eeq where we introduced \beq \Ham = \frac12 \omega^2 +V(\theta),
\quad V(\theta) = \frac12{\sin^2
  \theta}  +{\sigma p_\perp \theta}.
\label{DEFINEHAMILTONIAN}
\eeq At the next order, the effects of finite $\alpha$ and $\Lambda$
appear in the first-derivative term in \eqref{reducedODE}, while the
other forcing term is still higher order. The behavior of
\eqref{reducedODE} is therefore that of a weakly damped Hamiltonian
system with Hamiltonian $\Ham$, with the effects of $\alpha$ and
$\sigma$ serving to slowly change the value of $\Ham$ as the system
evolves. Thus, we now employ the technique of orbit-averaging to
reduce the problem further to the discrete dynamics of $\Ham(t)$,
where the discrete time-steps are equal (to the leading order) to
half-periods of the underlying Hamiltonian dynamics (which thus vary
with
$\Ham$). 

Let us first compute the continuous-in-time dynamics of $\Ham$. From \eqref{DEFINEHAMILTONIAN}, 
\beq
\dot{\Ham} = \omega(\dot \omega + V'(\theta)),
\eeq
which vanishes to leading order. At the next order, from \eqref{reducedODE}, one has 
\beq
\dot{\Ham} = -\frac{\omega^2}{\sqrt\Lambda}(\alpha + 2 \sigma \Lambda \cos\theta).
\label{DYNAMICSOFHAM}
\eeq

We now seek to average this dynamics over the Hamiltonian orbits. The general nature of the Hamiltonian orbits is either oscillations around a local minimum of $V(\theta)$ (limit cycles) or persistent precessions. If the local minimum of $V$ is close to an even multiple of $\pi$, $\Ham$ cannot increase, while if it is close to an odd multiple then $\Ham$ can increase if $\sigma$ is large enough. The switching process involves moving from the oscillatory orbits close to one of these odd minima, up the energy landscape, then jumping to oscillatory orbits around the neighboring even minimum, and decreasing in energy towards the new local fixed point.

We focus first on the oscillatory orbits. We may define their half-periods as
\beq
T(\Ham) = \int_{\theta^*_-}^{\theta^*_+}\frac{\d\theta}{\dot{\theta}},
\label{PERIODOSCILLATORY}
\eeq
where $\theta_-^*$ and $\theta_+^*$ are the roots of the equation $V(\theta)=\Ham$ to the left and right of the local minimum of $V(\theta)$ about which $\theta(t)$ oscillates.  To compute this integral, we assume that ${\theta(t)}$ follows the Hamiltonian trajectory:
\beq
\dot{\theta} = \pm \sqrt{2(\Ham - V(\theta))}.
\label{traj}
\eeq We then define the half-period average of a function $f(\theta(t))$ as \beq
\avg{f} = \frac{1}{T(\Ham)}
\int_{\theta^*_-}^{\theta^*_+}\frac{f(\theta)
  \d\theta}{\sqrt{2(\Ham -
    V(\theta))}}, \eeq which agrees with the time average over
  half-period to the leading order. Note that this formula applies
  irrespectively of whether the
  trajectory connects $\theta^*_-$ to $\theta^*_+$ or $\theta^*_+$ to
  $\theta^*_-$. Applying this averaging to $\dot{\Ham}$, we then have
  \beq \avg{\dot{\Ham}} = -\frac{1}{
  T(\Ham)}\int_{\theta^*_-}^{\theta^*_+}  \chi(\theta,\Ham) \d
\theta, 
\label{AVERAGEDDYNAMICS}
\eeq where we defined \beq \chi(\theta,\Ham) = \frac{\brackets{\alpha
    + 2 \sigma \Lambda \cos \theta}\sqrt{2(\Ham - V(\theta))}
}{\sqrt{\Lambda}}. \eeq

If the value of $\Ham$ is such that either of the roots $\theta^*_\pm$
no longer exist, this indicates that the system is now on a
precessional trajectory. In order to account for this, we can define
the period on a precessional trajectory instead as
\beq
T(\Ham) =  \int_{\theta_C - \pi}^{\theta_C}\frac{\d\theta}{\dot{\theta}},
\label{PERIODPRECESSIONAL}
\eeq where $\theta_C$ is a local maximum of $V(\theta)$. On the
precessional trajectories, we then have \beq \avg{\dot{\Ham}} = -
\frac{1}{ T(\Ham)}\int_{\theta_C-\pi}^{\theta_C} \chi(\theta,\Ham) \d
\theta. \eeq

In order to approximate the ODE solutions, we now decompose the
dynamics of  $\Ham$ into half-period time intervals. 
We thus take, at the $n$'th timestep, $\Ham_n = \Ham(t_n)$, $t_{n+1} =
t_n + T(\Ham_n)$ and
\beq
\Ham_{n+1} = \Ham_n - \int_{\theta^*_-(\Ham_n)}^{\theta^*_+(\Ham_n)} \chi(\theta,\Ham_n) \d \theta,
\label{discretemap}
\eeq if $\Ham_n$ corresponds to a limit cycle trajectory. The same
discrete map applies to precessional trajectories, but with the
integration limits replaced with $\theta_C - \pi$ and $\theta_C$,
respectively.


\subsection{Modelling switching with discrete map}

In order to model switching starting from inside a well of $V(\theta)$, we can iterate the discrete map above, starting from an initial energy $\Ham_0$. We choose $\Ham_0$ by choosing a static initial condition $\theta(0)=\theta_0$ close to an odd multiple of $\pi$ (let us assume without loss of generality that we are close to $\pi$), and computing $\Ham_0 = V(\theta_0)$. 

On the oscillatory trajectories, the discrete map then predicts the
maximum amplitudes of oscillation ($\theta^*_\pm(\Ham_n)$) at each
timestep, by locally solving $\Ham_n = V(\theta)$ for each $n$. After
some number of iterations, the trajectory will escape the local
potential well, and one or both roots of $\Ham_n = V(\theta)$ will not
exist. Due to the positive average slope of $V(\theta)$ the most
likely direction for a trajectory to escape the potential well is
$\dot{\theta}<0$ (`downhill'). Assuming this to be the case, at some
timestep $t_N$, it will occur that the equation
$\Ham_N = V(\theta)$ has only one root
$\theta = \theta^*_+ > \pi$, implying that the trajectory has
escaped the potential well, and will proceed on a precessional
trajectory in a negative direction past $\theta=\pi/2$ towards
$\theta=0$.

\begin{figure*}[t]
  \begin{center}
    \includegraphics[width=.9\textwidth]{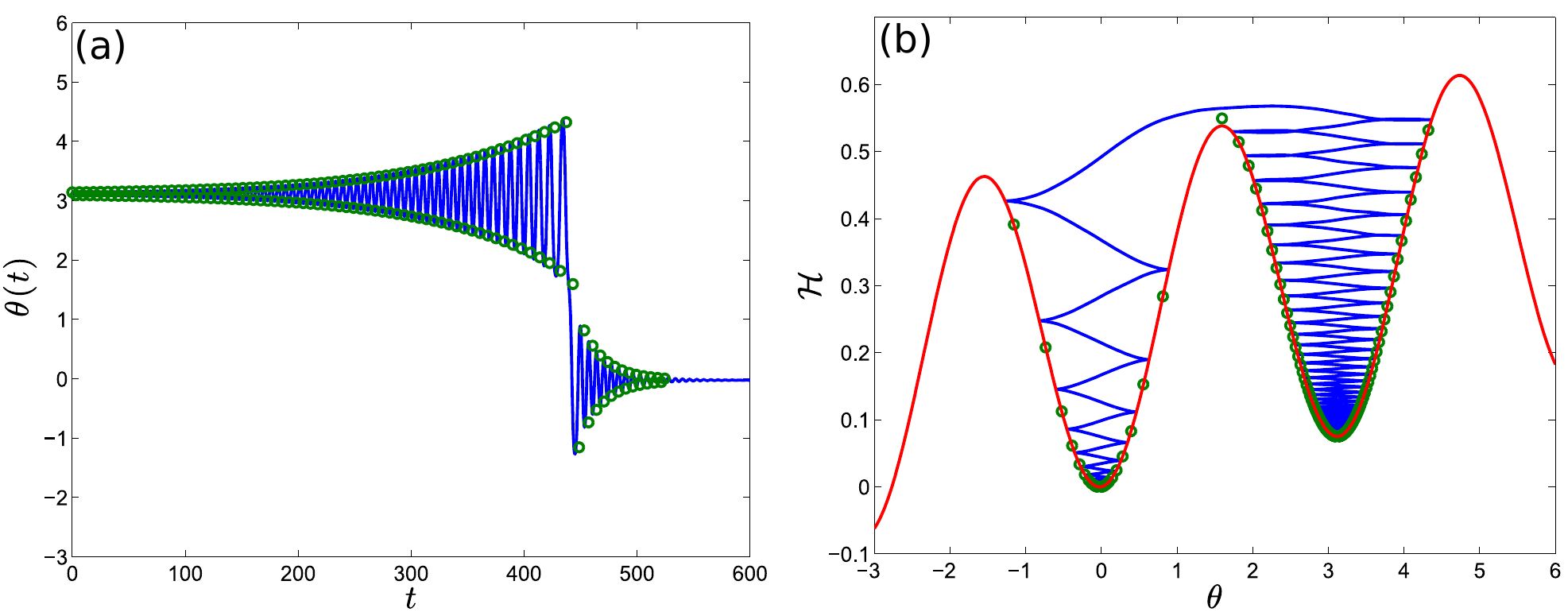}
    \caption{Switching solution (blue line) and its discrete
      approximation (green circles). Parameters: $\alpha=0.01$,
      $\Lambda=0.1$, $p_\perp = 0.3$, $\sigma=0.08$. Panel (a) shows
      the solution $\theta(t)$, and panel (b) shows the trajectory for
      this solution in the $\Ham-\theta$ plane.
      The red line in (b) shows $V(\theta)$.}
      \label{averaged-switching} 
    \end{center} 
  \end{figure*}

To distinguish whether a trajectory results in switching or precession, we then perform a single half-period step on the precessional orbit from $\theta_C$ to $\theta_C - \pi$, and check whether $\Ham < V(\theta_C-\pi)$: if this is the case, the trajectory moves back to the oscillatory orbits around the well close to $\theta=0$, and decreases in energy towards the fixed point near $\theta=0$, representing switching. If however $\Ham > V(\theta_C-\pi)$ after the precessional half-period, the solution will continue to precess. 

In Fig. \ref{averaged-switching} below, we display the result of such
an iterated application of the discrete map, for the same parameters
as the switching solution given in Fig. \ref{odesols}(c). In Fig.
\ref{averaged-switching}(a), the continuous curve represents the
solution to \eqref{reducedODE}, and the points are the predicted peaks
of the oscillations, from the discrete map \eqref{discretemap}. Fig.
\ref{averaged-switching}(b) shows the energy of the same solution as a
function of $\theta$.
Again the blue curve gives $\Ham(t)$ for the ODE solution, the green points are the prediction of the iterated discrete map, and the red curve is $V(\theta)$.   The discrete map predicts the switching behavior quite well, only suffering some error near the switching event, when the change of $\Ham$ is significant on a single period.

\subsection{Modelling precession}

\begin{figure*}[t]
\begin{center}
\includegraphics[width=.9\textwidth]{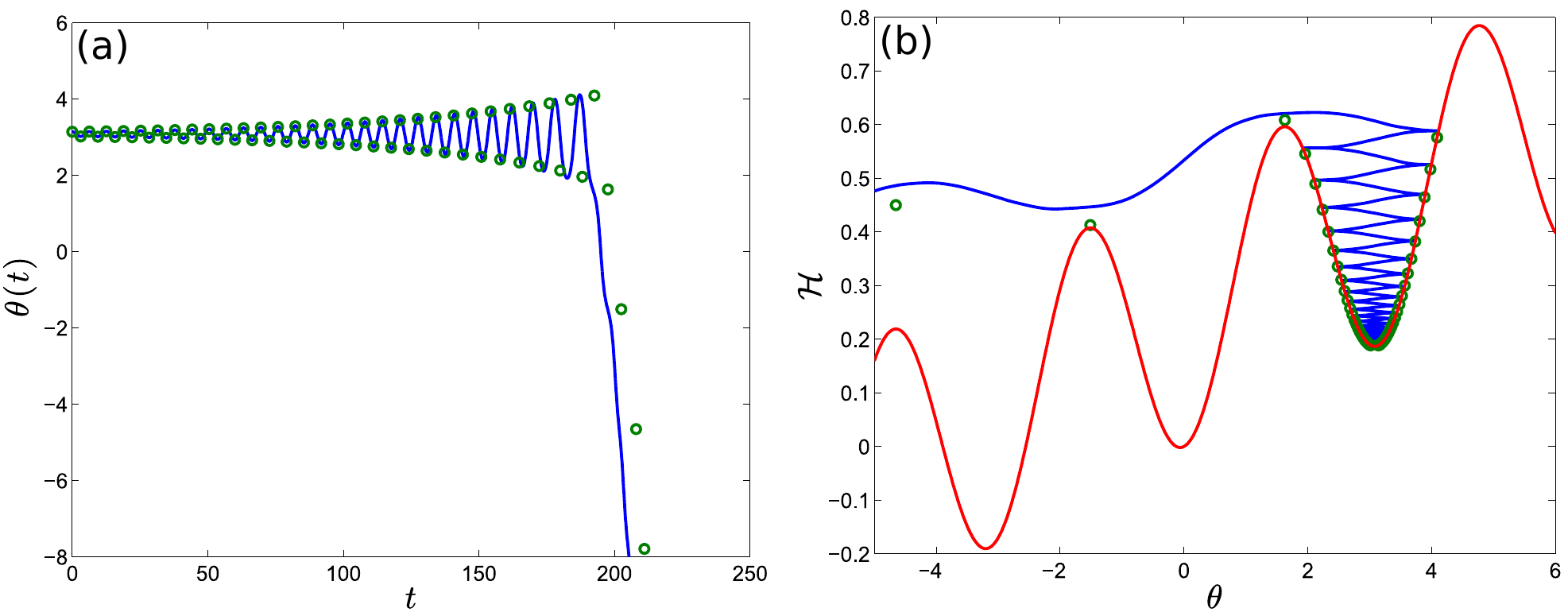}
\caption{Precessing solution (blue line) and its discrete
  approximation (green circles). Parameters: $\alpha=0.01$,
  $\Lambda=0.1$, $p_\perp = 0.6$, $\sigma=0.1$. Panel (a) shows the
  solution $\theta(t)$, and panel (b)  shows the trajectory for
  this solution in the $\Ham-\theta$ plane. The red line in
    (b) shows $V(\theta)$.}
\label{averaged-precession}
\end{center}
\end{figure*}

Here we apply the discrete map to a precessional solution---one in
which the trajectory, once it escapes the potential well near $\pi$,
does not get trapped in the next well, and continues to rotate. Fig.
\ref{averaged-precession}(a) below displays such a solution
$\theta(t)$ and its discrete approximation, and Fig.
\ref{averaged-precession}(b) displays the energy of the same solution.
Again, the prediction of the discrete map is excellent.

\section{Transitions in trajectories}

In this section we seek to understand the transitions between the trapping, switching, and precessional regimes as the current parameters $\sigma$ and $p_\perp$ are varied. 
\subsection{Escape Transition}
Firstly, let us consider the transition from states which are trapped
in a single potential well, such as those in Figs. \ref{odesols}(a,b),
to states which can escape and either switch or precess. Effectively,
the absolute threshold for this transition is for the value of $\Ham$
to be able to increase for some value $\theta$ close to the minimum of
$V(\theta)$ near $\pi$. Thus, we consider the equation of motion
\eqref{DYNAMICSOFHAM} for $\Ham$, and wish to find parameter values
such that $\dot{\Ham} > 0$ for some $\theta$ near $\pi$. This
  requires that \beq \frac{\omega^2}{\sqrt\Lambda}(\alpha + 2 \sigma
\Lambda \cos \theta) < 0. \eeq Assuming that $\omega \neq 0$, we can
see that the optimal value of $\theta$ to hope to satisfy this
condition is $\theta=\pi$, yielding a theoretical minimum
$\sigma = \sigma_s$
  for the
  dimensionless current density for motion to be possible, with
\beq \sigma_s = \frac{\alpha}{2\Lambda}. \eeq This is similar to
the critical switching currents derived in previous work
\cite{Pinna2013}. We then require $\sigma > \sigma_s$ for the
possibility of switching or precession. Note that this estimate is
independent of the value of $p_\perp$.

\subsection{Switching--Precessing Transition}
We now consider the transition from switching to precessional states.
This is rather sensitive and there is not in general a sharp
transition from switching to precession. It is due to the fact
that for certain parameters, the path that the trajectory takes once
it escapes the potential well depends on how much energy it has as it
does so. In fact, for a fixed $\alpha, \Lambda$, and values of
$\sigma > \sigma_s$
we can separate the $(\sigma, p_\perp)$-parameter space into three
regions: (i) after escaping the initial well, the trajectory
always falls into the next well, and thus switches; (ii) after
escaping, the trajectory may either switch or precess depending on its
energy as it does so (and thus depending on its initial condition);
(iii) after escaping, the trajectory completely passes the next well,
and thus begins to
precess. 

We can determine in which region of the parameter space a given point $(\sigma,p_\perp)$ lies by studying the discrete map \eqref{discretemap} close to the peaks of $V(\theta)$. Assume that the trajectory begins at $\theta(0) = \pi$, and is thus initially in the potential well spanning the interval $\pi/2 \leq \theta \leq 3\pi/2$. Denote by $\theta_C$ the point close to $\theta=\pi/2$ at which $V(\theta)$ has a local maximum. 
It is simple to compute
\beq
\theta_C = \frac{\pi}{2} + \frac{1}{2} \sin^{-1}(2\sigma p_\perp).
\eeq
Moreover, it is easy to see that all other local maxima of $V(\theta)$ are given by $\theta = \theta_C + k\pi$, for $k\in \mathbb{Z}$. 

We now consider trajectories which escape the initial well by crossing $\theta_C$. 
These trajectories have, for some value of the timestep $n$ while still confined in the initial well, an energy value $\Ham_n$ in the range
\beq
\Ham_{\text{\scriptsize{trap}}} < \Ham_n < V(\theta_C+\pi),
\label{range}
\eeq 
where we define $\Ham_{\text{\scriptsize{trap}}}$ to be the value of $\Ham_n$ such that the discrete map \eqref{discretemap} gives $\Ham_{n+1} = V(\theta_C)$. We thus have $\Ham_{n+1} > V(\theta_C)$. In order to check whether the trajectory switches or precesses, we then compute $\Ham_{n+2}$ and compare it to $V(\theta_C-\pi)$. We may then classify the trajectories as switching if $\Ham_{n+2} - V(\theta_C-\pi) < 0$, and precessional if $\Ham_{n+2} - V(\theta_C-\pi) > 0$.

\begin{figure}[b]
\begin{center}
\includegraphics[width=.45\textwidth,trim=40 0 0 20, clip=true]{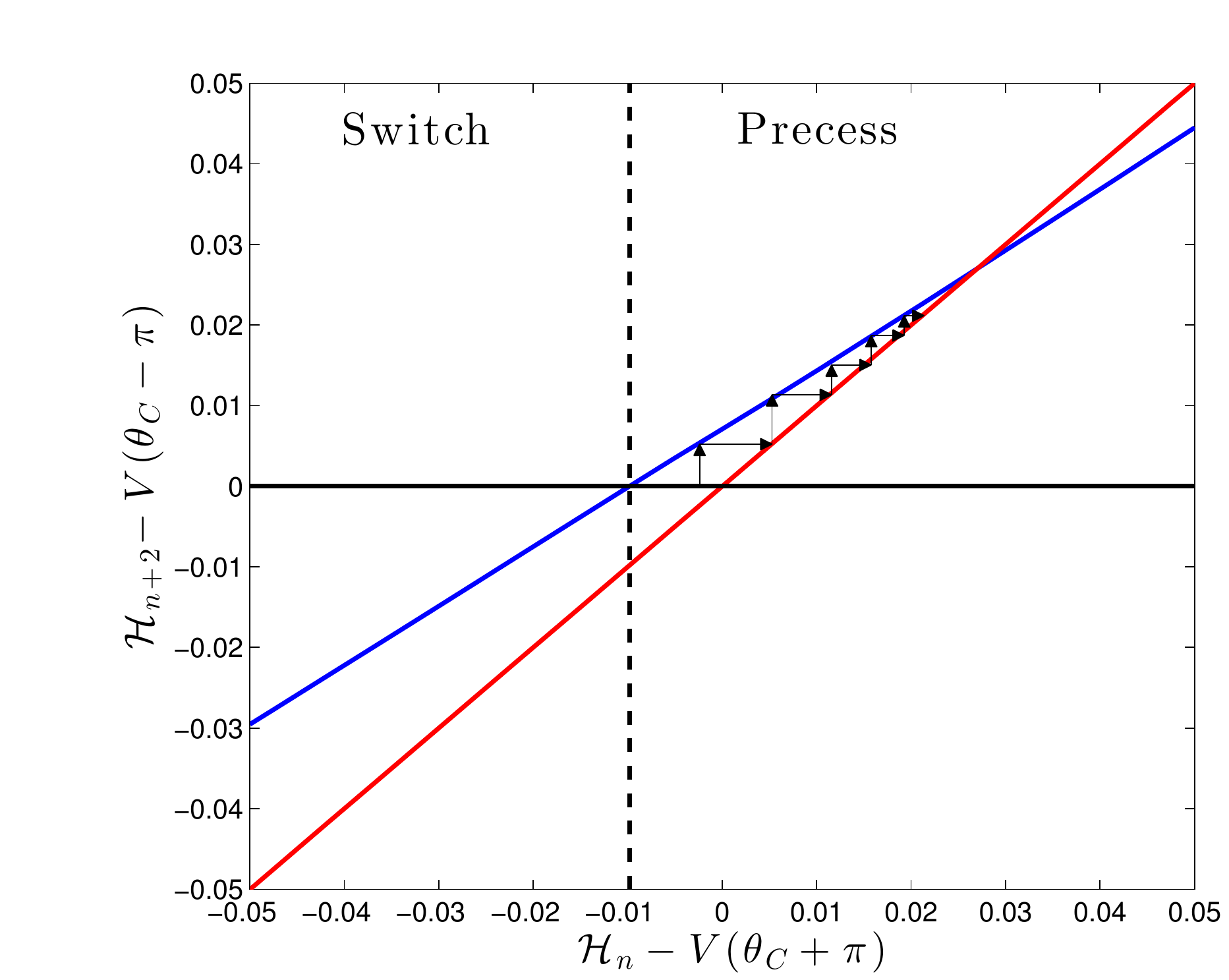}
\caption{Precession vs switching prediction from the discrete map. Parameters: $\alpha=0.01$, $\Lambda=0.1$, $p_\perp = 0.35$, $\sigma=0.08$. Values of $\Ham_n-V(\theta_C+\pi)$ to the left of the dashed line switch after the next period, the trajectory becoming trapped in the well around $\theta=0$. Values to the right begin to precess, and converge to a precessional fixed point of the discrete map.}
\label{prec_v_switch}
\end{center}
\end{figure}

Figure \ref{prec_v_switch} displays a plot of
$\Ham_n - V(\theta_C+\pi)$ against $\Ham_{n+2} - V(\theta_C-\pi)$. The
blue line shows the result of applying the discrete map, while the red
line is the identity line. Values of $\Ham_n-V(\theta_C+\pi)$ which
are inside the range specified in \eqref{range} are thus on the
negative $x$-axis here. We can classify switching trajectories as
those for which the blue line lies below the $x$-axis, and precessing
trajectories as those which lie above. In Fig. \ref{prec_v_switch},
the parameters are such that both of these trajectory types are
possible, depending on the initial value of $\Ham_n$, and thus this
set of parameters are in region (ii) of the parameter space. We note
that, since the curve of blue points and the identity line intersect
for some large enough value of $\Ham$, this figure implies that if the
trajectory has enough energy to begin precessing, then after several
precessions the trajectory will converge to one which conserves energy
on average over a precessional period (indicated by the arrows). In
region (i) of the parameter space, the portion of the blue line for
$\Ham_n - V(\theta_C+\pi) < 0$ would have
$\Ham_{n+2} - V(\theta_C-\pi) <0$, while in region (iii), they would
all have $\Ham_{n+2} - V(\theta_C-\pi) >0$.

We can classify the parameter regimes for which switching in the opposite direction (i.e. $\theta$ switches from $\pi$ to $2\pi$) is possible in a similar way. It is not possible to have a precessional trajectory moving in this  direction ($\dot\theta >0$), though.

We may then predict, for a given point $(\sigma, p_\perp)$ in
parameter space, by computing relations similar to that in Fig.
\ref{prec_v_switch}, which region that point is in, and thus generate
a theoretical phase diagram.

In Fig. \ref{phase} below, we display the phase diagram in the
$(\sigma, p_\perp)$-parameter space, showing the end results of
solving the ODE \eqref{reducedODE} as a background color, together
with predictions of the bounding curves of the three regions of the
space, made using the procedure described above. The predictions of
the discrete map, while not perfect, are quite good, and provide
useful estimates on the different regions of parameter space. In
particular, we note that the region where downhill switching reliably
occurs (the portion of region (i) above the dashed black line) is
estimated quite well. We would also note that we would expect the
predictions of the discrete map to improve if the values of $\Lambda$
and $\alpha$ were decreased.

\begin{figure}[h]
\begin{center}
\includegraphics[width=.45\textwidth, trim=20 0 0 20,clip=true]{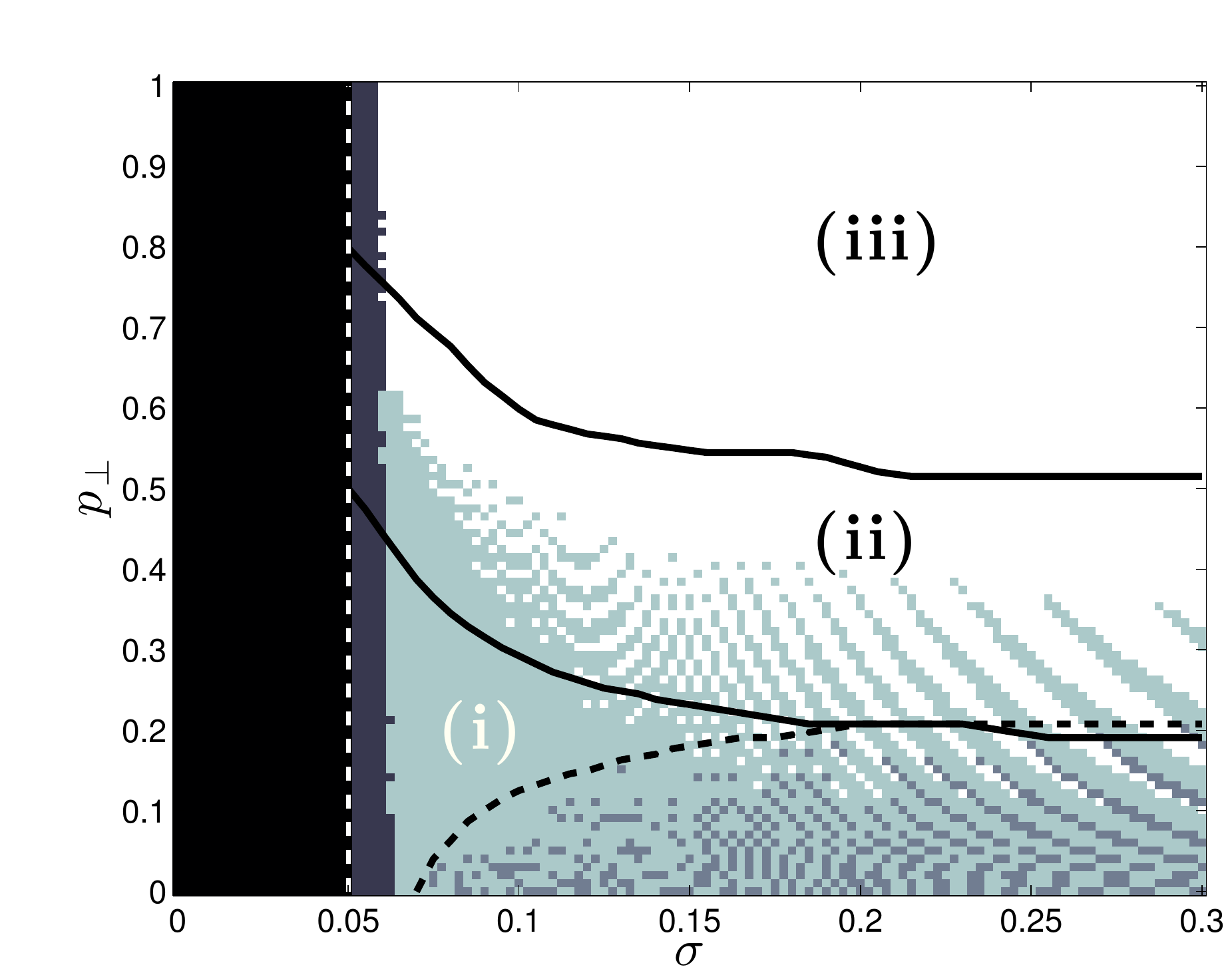} 
\caption{Macrospin solution phase diagram: $\alpha=0.01, \Lambda=0.1$.
  The background color indicates the result of solving the ODE
  \eqref{reducedODE} with initial condition \eqref{ICMacro}: the dark
  region to the left of the figure indicates solutions which do not
  escape their initial potential well, and the vertical dashed white
  line shows the computed value of the minimum current required to
  escape, $\sigma_s = \alpha/(2\Lambda)$. The black band represents
  solutions which decay, like in Fig. \ref{odesols}(a),
  while the dark grey band represents solutions like in
  Fig. \ref{odesols}(b). In the rest of the figure, the green points
  indicate switching in the negative direction like in 
  Fig. \ref{odesols}(c), grey indicate switching in the positive
  direction, and white indicates precession like in 
  Fig. \ref{odesols}(d).
  The solid black curves are the predictions of boundaries of the
  regions (as indicated in the figure) by using the discrete map, and
  the dashed line is the prediction of the boundary below which
  switching in the positive direction is possible. }
\label{phase}
\end{center}
\end{figure}

\section{Discussion}

We have derived  an underdamped PDE model for
magnetization dynamics in thin films subject to perpendicular applied
spin-polarized currents, valid in the asymptotic regime of small
$\alpha$ and $\Lambda$, corresponding to weak damping and strong
penalty for out-of-plane magnetizations. We have examined the
predictions of this model applied to the case of an elliptical
film under a macrospin approximation by using an orbit-averaging
approach. We found that they qualitatively agree quite well with
previous simulations using full LLGS dynamics \cite{ChavesKent15}.

The benefits of our reduced model are that they should faithfully
reproduce the oscillatory nature of the in-plane magnetization
dynamics, reducing computational expense compared to full
  micromagnetic simulations. In particular, in
  sufficiently small and thin magnetic
  elements the problem further reduces to a single
second-order scalar equation.

The orbit-averaging approach taken here enables the investigation of
the transition from switching to precession via a simple discrete
dynamical system. The regions in parameter space where either
switching or precession are predicted, as well as an intermediate
region where the end result depends sensitively on initial conditions.
It may be possible to further probe this region by including either
spatial variations in the magnetization (which, in an earlier study
\cite{ChavesKent15} were observed to simply `slow down' the dynamics
and increase the size of the switching region), or by including
thermal noise, which could result in instead a phase diagram
predicting switching probabilities at a given temperature, or both.

\section*{ACKNOWLEDGMENTS} 
  Research at NJIT was supported in part by NSF via Grant No.
    DMS-1313687. Research at NYU was
  supported in part by NSF via Grant No.
  DMR-1309202.

\end{document}